\documentclass[12pt]{article}

\usepackage{amssymb}
\usepackage{amsbsy}
\usepackage{amsmath}

\newcommand{\be}{\begin{equation}} \newcommand{\ee}{\end{equation}}
\newcommand{\bea}{\begin{eqnarray}} \newcommand{\eea}{\end{eqnarray}}
\newcommand{\el}{\nonumber \\}
\newcommand{\re}[1]{(\ref{#1})}
\newcommand{\pat}{\partial}
\newcommand{\adot}{\dot{a}} \newcommand{\addot}{\ddot{a}}
\newcommand{\bdot}{\dot{b}} \newcommand{\bddot}{\ddot{b}}
\newcommand{\ndot}{\dot{n}}

\newcommand{\phidot}{\dot{\phi}}
\newcommand{\phiddot}{\ddot{\phi}}
\newcommand{\rhodot}{\dot{\rho}}
\newcommand{\pdot}{\dot{p}}
\newcommand{\fdot}{\dot{f}}

\begin{document}

\begin{titlepage}
\begin{flushleft}
       \hfill                      {\tt hep-th/0106282}\\ \hfill
       HIP-2001-28/TH \\ \hfill            June 29, 2001\\
\end{flushleft}
\vspace*{3mm}
\begin{center}
{\Large {\bf Hubble Law and Brane Matter after Ekpyrosis}\\}
\vspace*{12mm} {\large Kari Enqvist\footnote{E-mail:
kari.enqvist@helsinki.fi}, Esko Keski-Vakkuri\footnote{E-mail:
esko.keski-vakkuri@hip.fi} and Syksy
R\"{a}s\"{a}nen\footnote{E-mail: syksy.rasanen@helsinki.fi}\\}

\vspace{5mm}

{\em {}$^{1,2}$Helsinki Institute of Physics \\ P.O. Box 64,
FIN-00014 University of Helsinki, Finland }

\vspace{3mm}

and

\vspace{3mm}

{\em {}$^{1,3}$Department of Physics \\ P.O. Box 64, FIN-00014
University of Helsinki, Finland}

\vspace*{10mm}
\end{center}

\begin{abstract} \noindent

We study brane matter in the ekpyrotic scenario and observe that in order to
obtain standard gravity on the visible  brane, the tension of the visible
brane should be positive. If the sizes of both the fifth dimension and the
Calabi-Yau threefold are fixed, the Israel junction conditions do not allow
time-dependent brane matter. Relaxing this constraint, it is possible to
obtain approximately standard cosmology on the visible brane, with small
corrections due to possible time-dependence of the Calabi-Yau threefold.

\end{abstract}

\end{titlepage}

\baselineskip16pt
\setcounter{footnote}{0}

\section{Introduction}

Recently, a novel cosmological scenario based on
colliding branes was proposed \cite{Khoury:2001a}. According to
\cite{Khoury:2001a}, this scenario, entitled ``the ekpyrotic
universe'', is based on the Ho\v{r}ava-Witten model of M-theory on
a $S^1/Z_2$ orbifold \cite{HW}, compactified to five dimensions on
a Calabi-Yau threefold. The five-dimensional spacetime is bounded
by two orbifold planes, one corresponding to our universe (the
visible brane), and the other one to a hidden universe. Between the boundary
branes, a third brane travels across the bulk spacetime, eventually colliding
with and dissolving into the visible brane. The collision, called
``ekpyrosis'', heats up the visible brane, providing initial conditions for
the hot Big Bang. This model was presented as an alternative to the
inflationary scenario, resolving the horizon, flatness, and
monopole puzzles without involving superluminal expansion.
It was even suggested that the model gives a prediction for a
strongly blue gravitational wave spectrum, hoped to be
detectable in future experiments.

This is an interesting proposal; debate and more detailed examinations have
already begun \cite{Kallosh:2001, Khoury:2001b, Donagi:2001, Lyth:2001pf,
KKLT}. While many of the key issues, such as the initial
configuration of the branes, the dynamics of the bulk brane, and
the details of the collision of the bulk brane with the visible
brane involve a detailed understanding of the physics of
string/M-theory, there are other features which are more tractable
and can be studied in the framework of the effective field theory limit of
five-dimensional heterotic M-theory \cite{Lukas:1998a, Lukas:1998b}.
In this paper, we will investigate the cosmology on the visible
brane after the collision in the effective field theory context.

One of the debated issues is whether the visible brane should
have positive or negative tension \cite{Kallosh:2001,
Khoury:2001b, Donagi:2001, KKLT}. We propose that a simple way to
test that is to examine the Hubble law in the post-ekpyrosis
era. In the context of the related Randall-Sundrum model \cite{RanSun},
the Hubble law has been extensively investigated, and it is known
that obtaining (nearly) standard cosmology requires that\footnote{Or is at
least easiest when; see \cite{Kanti} for an alternative, which however is not
realised in the ekpyrotic scenario.} the visible brane has positive tension
\cite{Csaki:1999a, Cline:1999, Shiromizu:1999, Binetruy:1999b}. Another
feature known from investigations of the Randall-Sundrum model is that the
Israel junction conditions do not necessarily allow generic brane matter
\cite{Mohapatra:2000, Lesgourges:2000, Grinstein:2000, Enqvist:2000}. These
issues are relevant in the ekpyrotic scenario as well -- do the junction
conditions allow generic time-dependent matter on the visible brane, and can a
(nearly) standard Hubble law be recovered?

As we discuss in sections 2.2 and 2.3, both of the above requirements
may be met, but at a price. The junction conditions allow time-dependent matter
on the visible brane only if one is willing to allow time
evolution for the breathing modulus of the Calabi-Yau threefold
or the proper distance of the orbifold planes (or possibly
some other moduli). An almost standard Hubble law can be recovered if
the tension of the visible brane is positive, as in the Randall-Sundrum model. 
The required cancellations originate from the (near) BPS property of the brane
configuration, in the spirit of \cite{Cvetic}.

The requirement of positive tension for the visible brane does not violate the
principles of the ekpyrotic scenario, but helps to narrow the range of
possibilities for its realization (see \cite{Kallosh:2001, Khoury:2001b} for
discussion). The result on brane matter calls for additional analysis of the
dynamics of the effective field theory with time-dependent moduli.

\section{Brane cosmology after ekpyrosis}

\subsection{The set-up}

\paragraph{The action and the metric.}

The five-dimensional effective action of heterotic M-theory is quite
complicated \cite{Lukas:1998a, Lukas:1998b}. In \cite{Khoury:2001a}, a
simplified action used as an example of the realization of the ekpyrotic
scenario was given as\footnote{In \cite{Khoury:2001a}, there seems to be a
factor $\sqrt{2/3}$ missing in front of ${\cal A}_{ABCD}$. This is explained in \cite{KKLT}.}
\bea \label{action1}
  & & \!\!\!\!\!\! \!\!\!\!\!\! S_{\textrm{het}} \,=\, \frac{M_5^3}{2}\int_{{\cal M}_5} d^5 x \sqrt{-g}\left(R-\frac{1}{2}\pat_A\phi\,\pat^A\phi-\frac{1}{5!}e^{2\phi}{\cal F}_{ABCDE}{\cal F}^{ABCDE}\right) \el
  & & \!\!\!\!\!\! \!\!\!\!\!\! \!\!\!\!\!\! - \sum_{i=1}^2 3\alpha_i M_5^3\int_{{\cal M}_4^{(i)}}\!\! d^4\xi_{(i)}\left(\sqrt{-h_{(i)}}e^{-\phi} - \frac{1}{4!}\epsilon^{\mu\nu\kappa\lambda}\sqrt{\frac{2}{3}}{\cal A}_{ABCD}\pat_{\mu}X^A_{(i)}\pat_{\nu}X^B_{(i)}\pat_{\kappa}X^C_{(i)}\pat_{\lambda}X^D_{(i)}\right) ,
\eea

\noindent where $M_5$ is the five-dimensional Planck mass, $R$ is the
five-dimensional Ricci scalar, $\phi$ is the breathing modulus of the
Calabi-Yau threefold and ${\cal A}_{ABCD}$ is a four-form gauge field with
field strength ${\cal F}=d{\cal A}$. The Latin indices run from 0 to 4 and the
Greek indices run from 0 to 3. The spacetime is a five-dimensional manifold
${\cal M}_5= {\cal M}_4\times S_1/Z_2$ with coordinates $x^A$. The
four-dimensional manifolds ${\cal M}_4^{(i)}$, $i=1,2$, are the
end-of-the-world orbifold planes, called the visible and hidden branes
respectively, with internal coordinates $\xi^{\mu}_{(i)}$ and tensions
$\alpha_i M_5^3$. The tensor $g_{AB}$ is the metric on ${\cal M}_5$,
$h^{(i)}_{\mu\nu}$ is the induced metric on ${\cal M}_4^{(i)}$, and $g$ and
$h_{(i)}$ are their determinants. The functions $X^A_{(i)}(\xi_{(i)}^{\mu})$
are the coordinates in ${\cal M}_5$ of a point on ${\cal M}_4^{(i)}$ with
coordinates $\xi_{(i)}^{\mu}$. In other words, $X^A_{(i)}(\xi_{(i)}^{\mu})$
encode the embedding of the branes into ${\cal M}_5$. The symbol
$\epsilon^{\mu\nu\kappa\lambda}$ is the antisymmetrization operator,
with $\epsilon^{0123}=1$.

We are interested in post-ekpyrosis cosmology, so the bulk brane present in
\cite{Khoury:2001a} has already collided with and dissolved into the visible
brane. We will not consider ekpyrosis (the collision); following
\cite{Khoury:2001a} we assume that matter is created on the visible brane
and so add to the action the term
\bea \label{matteraction}
  S_{\textrm{matter}} = \int_{{\cal M}_4^{(1)}} d^4\xi_{(1)}\sqrt{-h_{(1)}}{\cal L}_{\textrm{matter}} \ .
\eea

We will neglect the possibility of bending of the branes and take their
embedding to be the same as in the BPS state,
\bea \label{embed}
  X^A_{(i)}(\xi_{(i)}^{\mu}) = (x^0,x^1,x^2,x^3,y_i) \ ,
\eea

\noindent with $y_1=0$, $y_2=R$. We will consider the following metric
($t\equiv x^0$):\footnote{Note that our $n$ and $a$ are not the same lapse
function and expansion factor as the ones denoted by $n$ and $a$ in
\cite{Khoury:2001a}.}
\bea \label{genmetric}
  ds^2 = -n(t,y)^2 dt^2 + a(t,y)^2 \sum^3_{k=1}(dx^k)^2 + b(t,y)^2 dy^2 \ .
\eea

Homogeneity and isotropy with respect to the spatial directions
parallel to the branes would allow a $dt dy$--crossterm. However,
following the custom in the literature, we will neglect the crossterm.
We will denote partial derivatives with respect to $t$ and $y$ by dots and
primes. Given the embedding \re{embed}, the total action reads
\bea \label{action2}
  S &=& M_5^3\int_{{\cal M}_5} d^5 x \sqrt{-g}\bigg[
\frac{1}{2} R-\frac{1}{4}\pat_A\phi\,\pat^A\phi-\frac{1}{2}\frac{1}{5!}e^{2\phi}{\cal F}_{ABCDE}{\cal F}^{ABCDE} \el
  & & + \sum_{i=1}^2\delta(y-y_i) b^{-1} \left(-3\,\alpha_i e^{-\phi} + (-h_{(i)})^{-1/2}\sqrt{6}\,\alpha_i{\cal A}_{0123} + \delta_{i1} M_5^{-3}{\cal L}_{\textrm{matter}}\right)\bigg] \ .
\eea

Because of the $S_1/Z_2$ orbifold structure of ${\cal M}_5$, the first
$y$-derivatives of the metric and of the fields $\phi$ and ${\cal A}_{ABCD}$
are discontinuous at the brane locations, and their second derivatives
contain delta functions related to the discontinuities. For example, the
delta function part of $a''$ is \cite{Binetruy:1999a}
\bea \label{deltacont}
  a''\big|_\delta &=& \sum_{i=1,2}\delta(y-y_i)[a'] \el
  &=& \sum_{i=1,2}\delta(y-y_i)(-1)^{i+1} 2\, a'_c \ ,
\eea

\noindent where $[a']$ is the discontinuity of $a'$,
$[a'(y)]:=\lim_{\varepsilon\rightarrow0}
(a'(y+\varepsilon)-a'(y-\varepsilon))$, and $a'_c$ is the continuous part of
$a'$. An identical relation holds for $n$, $\phi$, and ${\cal A}_{ABCD}$.

\paragraph{The equations of motion.}

From the action \re{action2} we obtain the following field equations for
$\phi$ and ${\cal A}_{ABCD}$:
\bea \label{eom1}
  \square\phi-\frac{2}{5!}e^{2\phi}{\cal F}_{ABCDE}{\cal F}^{ABCDE}
+ \sum_{i=1}^2\delta(y-y_i) b^{-1}6\,\alpha_i e^{-\phi} &=& 0 \\
  D_M (e^{2\phi}{\cal F}^{MABCD}) + \delta_0^{[A}\delta_1^{\ B}\delta_2^{\ C}\delta_3^{\ D]} \sum_{i=1}^2\delta(y-y_i) (-g)^{-1/2}\sqrt{6}\,\alpha_i &=& 0 \ ,
\eea

\noindent where $D_M$ is the covariant derivative. Writing the delta
function parts of the second $y$-derivative of $\phi$ and the first
$y$-derivative of ${\cal F}_{MABCD}$ explicitly, we have
\bea \label{eom2}
   -n^{-2} \left[\phiddot + \left(-\frac{\ndot}{n}+3\frac{\adot}{a}+\frac{\bdot}{b}\right)\phidot\right] \el
  + b^{-2} \left[\phi''_c + \left(\frac{n'_c}{n}+3\frac{a'_c}{a}-\frac{b'_c}{b}\right)\phi'_c\right] +\, 2\, n^{-2} a^{-6} b^{-2} e^{2\phi} ({\cal F}_{0123y\, c})^2 &=& 0 \el
  \pat_t (n^{-1} a^{-3} b^{-1} e^{2\phi}{\cal F}_{0123y\, c}) &=& 0 \el
  \pat_y (n^{-1} a^{-3} b^{-1} e^{2\phi}{\cal F}_{0123y\, c}) &=& 0 \el
  \delta(y-y_i) \big((-1)^{i+1}\phi'_c + 3\, b\,\alpha_i\, e^{-\phi}\big) &=& 0 \el
  \delta(y-y_i) \big(2 (-1)^i n^{-1} a^{-3} b^{-1} e^{2\phi}{\cal F}_{0123y\, c} + \sqrt{6}\, \alpha_i\big) &=& 0 \ .
\eea

In what follows, we suppress the subscript $c$ for convenience; it should be
clear from the context whether it is the function or its continuous part that
is meant. Solving for the field strength ${\cal F}_{0123y}$, we get
\bea \label{eom3}
  {\cal F}_{0123y} &=& -\sqrt{\frac{3}{2}}\, \alpha n a^3 b\, e^{-2\phi} \\
  \alpha_i &=& (-1)^i \alpha \\
  \label{eomphi} - n^{-2} \left[\phiddot + \left(-\frac{\ndot}{n}+3\frac{\adot}{a}+\frac{\bdot}{b}\right)\phidot\right] \el
  + b^{-2} \left[\phi'' + \left(\frac{n'}{n}+3\frac{a'}{a}-\frac{b'}{b}\right)\phi'\right] + 3\,\alpha^2 e^{-2\phi} &=& 0 \\
   \label{eomphib} \delta(y-y_i) (\phi'-3\, b\alpha\, e^{-\phi}) &=& 0 \ .
\eea

We follow the convention of \cite{Khoury:2001a}: the hidden brane has
tension $\alpha M_5^3$ and the visible brane has tension $-\alpha M_5^3$.
In \cite{Khoury:2001a}, $\alpha$ was positive so that the visible brane had
negative tension; we leave the sign undetermined for the time being.

\paragraph{The Einstein equation.}

In addition to the field equations of $\phi$ and ${\cal A}_{ABCD}$,
we must take into account the Einstein equation
\bea
  G_{A B} = \frac{1}{M_5^3}T_{A B} \ .
\eea

In component form, the Einstein equation from the action \re{action2} reads
\bea \label{einstein1}
  G^t_{\ t} &=& \frac{3}{b^2} \left[\frac{a''}{a}+\frac{a'}{a}\left(\frac{a'}{a}-\frac{b'}{b}\right)\right] -\frac{3}{n^2}\frac{\adot}{a}\left(\frac{\adot}{a}+\frac{\bdot}{b}\right) \el
  &=& -\frac{1}{4} n^{-2}\phidot^2 - \frac{1}{4} b^{-2}\phi'{}^2 - \frac{3}{4}\alpha^2 e^{-2\phi} - \frac{1}{M_5^3}\sum_{i=1}^2 \delta(y-y_i) b^{-1}\rho_{b(i)}  \el
  G^k_{\ k} &=& \frac{1}{b^2} \left[2\frac{a''}{a}+\frac{n''}{n}+\frac{a'}{a}
\left(\frac{a'}{a}+2\frac{n'}{n}\right)-\frac{b'}{b}\left(\frac{n'}{n}
+2\frac{a'}{a} \right) \right] \el
  &&-\frac{1}{n^2}\left[2\frac{\addot}{a}+\frac{\bddot}{b}+\frac{\adot}{a}
\left(\frac{\adot}{a}-2\frac{\ndot}{n}\right)
+\frac{\bdot}{b}\left(2\frac{\adot}{a}-\frac{\ndot}{n}\right)\right] \el
  &=& \frac{1}{4} n^{-2}\phidot^2-\frac{1}{4} b^{-2}\phi'{}^2-\frac{3}{4}\alpha^2 e^{-2\phi} + \frac{1}{M_5^3}\sum_{i=1}^2\delta(y-y_i) b^{-1} p_{b(i)} \el
  G^y_{\ y} &=& \frac{3}{b^2} \frac{a'}{a} \left( \frac{a'}{a} +\frac{n'}{n}\right) - \frac{3}{n^2}\left[\frac{\addot}{a} + \frac{\adot}{a} \left(
\frac{\adot}{a} -\frac{\ndot}{n} \right)\right] \el
  &=& \frac{1}{4} n^{-2}\phidot^2+\frac{1}{4} b^{-2}\phi'{}^2-\frac{3}{4}\alpha^2 e^{-2\phi} \el
  G_{ty} &=& 3\left( \frac{n'}{n}\frac{\adot}{a}+\frac{a'}{a}\frac{\bdot}{b}
-\frac{\adot'}{a} \right) = \frac{1}{2}\phidot\, \phi'  \ ,
\eea

\noindent where $\rho_{b(i)}$ and $p_{b(i)}$ are the energy density
and pressure of brane $i$:
\bea \label{rhoandp}
  \rho_{b(i)} &=& \delta_{i1}\rho_m + 3 M_5^3\alpha_i e^{-\phi} \el
  p_{b(i)} &=& \delta_{i1} p_m - 3 M_5^3\alpha_i e^{-\phi} \ .
\eea

The terms $\rho_m$ and $p_m$ are the contribution of brane matter
($\cal{L}_{\textrm{matter}}$). Note that under the assumption of homogeneity
and isotropy, the energy-momentum tensor of brane matter necessarily has the
ideal fluid form. The delta function part of \re{einstein1} reads
\bea \label{einsteindelta}
  3\frac{1}{b}\frac{a''}{a}\bigg|_\delta
&=& -\frac{1}{M_5^3} \sum_{i=1}^2 \delta(y-y_i) \rho_{b(i)} \el
  \frac{1}{b} \left(2\frac{a''}{a}+\frac{n''}{n}\right)\bigg|_\delta
&=& \frac{1}{M_5^3} \sum_{i=1}^2 \delta(y-y_i) p_{b(i)} \ .
\eea

Using \re{deltacont}, the above equations can be rewritten as
\bea \label{branematter}
  (-1)^{i+1} \frac{1}{b}\frac{a'}{a}\bigg|_{y=y_i}
&=& -\frac{1}{6 M_5^3}\rho_{b(i)} \el
  (-1)^{i+1} \frac{1}{b}\frac{n'}{n}\bigg|_{y=y_i}
&=& \frac{1}{6 M_5^3} (2\rho_{b(i)} + 3 p_{b(i)})  \ .
\eea

\subsection{No brane matter in a static configuration}

As the basis of the realization of the ekpyrotic scenario,
\cite{Khoury:2001a} used the metric of the unique Poincare invariant
BPS solution of the simplified action \re{action1},
\bea \label{BPSmetric}
  ds^2 = D(y) \big(-N^2 dt^2 + A^2 \sum^3_{k=1}(dx^k)^2\big) + B^2 D(y)^4 dy^2 \ ,
\eea

\noindent where $D(y)=\alpha y +C$ and $N, A, B$ and $C$ are constants. The
functions $\phi$ and ${\cal F}_{0123y}$ can be found in \cite{Khoury:2001a,
KKLT}. Even allowing for time-dependence in the parameters $N, A, B$ and $C$,
we see that $a'/a=n'/n$. From \re{branematter} it then follows that
$\rho_m+p_m=0$, which implies that $\rhodot_m=0$.\footnote{Brane matter
satisfies the ordinary four-dimensional conservation law, as we will see in the
next section.}

Even the time-dependent extension of the BPS metric does not allow
for time-dependent matter on the brane. Actually, it is a well-known feature
of the Randall-Sundrum model that a factorisable metric cannot
support time-dependent brane matter \cite{Mohapatra:2000, Lesgourges:2000}.
More generally, it is known that constraints on brane matter may arise in
configurations which are static, meaning that the branes are at rest and the
size of the fifth dimension is fixed ($\bdot=0$) \cite{Grinstein:2000,
Enqvist:2000}. We will now show that this feature is found
in the ekpyrotic model as well.

We will look for constraints on brane matter in static configurations.
By a ``static configuration'' we mean a stabilized fifth dimension
(in coordinate systems where the branes are at rest, $\bdot=0$) and
a stabilized Calabi-Yau threefold ($\phidot=0$). The motivation for the second
condition is that the breathing modulus $\phi$ is related to the
Newton's constant measured on the brane. Therefore, obtaining standard
four-dimensional gravity on the brane would seem to require $\phidot$ to be
quite small.\footnote{The relation between $\phi$ and Newton's constant will be
considered in the next section.}

For a static configuration, the equation of motion of $\phi$, \re{eomphi},
reads
\bea
  b^{-2} \left(\phi'' + \left(\frac{n'}{n}+3\frac{a'}{a}-\frac{b'}{b}\right)\phi'\right) + 3\,\alpha^2 e^{-2\phi} &=& 0 \ .
\eea

Since $\bdot=0$ and $\phidot=0$, the time derivative of the above equation
gives
\bea
  \phi'\pat_t\left(\frac{n'}{n}+3\frac{a'}{a}\right)=0 \ .
\eea

We now consider the location of the visible brane, $y=y_1$. The function
$\phi'$ cannot be zero at the branes, as we see from \re{eomphib}. Relating
$n'/n$ and $a'/a$ to the density and pressure of brane matter, using
\re{branematter}, we obtain the condition
\bea
  \rhodot_m - 3 \pdot_m = 0 \ .
\eea

Thus, the only matter allowed on the visible (or hidden) brane by the
equation of motion of $\phi$ is a combination of radiation and vacuum energy.
However, from the Einstein equation one can deduce that even radiation is
disallowed.\footnote{To see this, it is most straightforward to consider the
combination $G^t_{\ t}+3G^k_{\ k} -2G^y_{\ y}$ and use (21).} Hence a
static configuration cannot support time-dependent brane matter. One may ask
if turning on additional fields not present in the simplified action
\re{action2} could help the situation. However, additional fields can only
help if they couple to the breathing modulus $\phi$ in such a manner as to
contribute new time-dependent terms to its equation of motion.

This is our first result: in order to have time-dependent brane matter,
one must have $\bdot\neq0$ or $\phidot\neq0$. The time-dependence of brane
matter is related to the time-dependence of the size of the hidden dimensions.
Turning the argument around, time-dependent brane matter will cause either the
fifth dimension or the Calabi-Yau threefold to vary in tune with the
expansion of the visible universe.

In the limit of vanishing brane matter, we would expect to find a
time-independent solution such as the BPS solution of \cite{Khoury:2001a}.
Then not only the presence but also the amount of time-dependence in $b$ and
$\phi$ would seem to depend on brane matter. One may ask whether
it is possible to recover approximately standard cosmology on the brane at
all, or whether the time-dependence of the extra dimensions will always spoil
attempts to obtain the usual Friedmann-Robertson-Walker equations with brane
matter as the dominant source. If this is possible, it is then interesting to
ask whether the small deviations from standard cosmology will be observable.

\subsection{Cosmology on the brane}

\paragraph{Hubble law on the branes.}

We will now consider the cosmological evolution as seen by an observer
on the visible brane (the results would also apply to the hidden
brane, were it to contain matter). In this section, we assume that
it is possible to introduce generic time-dependent matter on the visible
brane (by a suitable generalization of the scenario, as discussed above),
and proceed to examine if standard cosmology can be recovered.

The topic of brane cosmology has been extensively studied in the context
of the Randall-Sundrum model [12-25],
to mention a few references, and some of the results of these investigations
are relevant also to the ekpyrotic model. Cosmology on the branes in the case
of a scalar field with a potential in the bulk and on the branes has been
studied in [26-29].

Bulk matter content (and more generally, bulk curvature) will in general
affect cosmology on the branes. Therefore, in order to obtain the expansion
laws on the brane, one would (in general) first have to solve the Einstein
equation (or at least the field equations) in the bulk. We will however
follow an easier route and analyze only the brane part of
the equations. This shorter road has two caveats to keep in mind.
First, in the brane equations there will appear bulk fields as well as a
radiation-like term \cite{Binetruy:1999b, Mohapatra:2000, Ida:1999,
Mukohyama:2000}, the magnitude of which cannot in general be fixed without
reference to the bulk solution. The second, more subtle, point is that there
is no guarantee that a solution of the brane equations is also a solution
of the full set of (bulk) equations.

We will use the general metric \re{genmetric}, and make no assumption about
the time-dependence of $b$ or $\phi$. We assume the visible brane to lie at a
fixed position, and with no loss of generality choose coordinates where the
lapse function $n(t,y_1)=1$, corresponding to cosmic time on the visible brane.
The $yy$-- and $ty$--components of the Einstein equation \re{einstein1} are
\bea \label{einstein2}
  \frac{1}{b^2} \frac{a'}{a} \left( \frac{a'}{a} +\frac{n'}{n}\right) - \frac{1}{n^2}\left[\frac{\addot}{a} + \frac{\adot}{a} \left(\frac{\adot}{a} -\frac{\ndot}{n} \right)\right]
  &=& \frac{1}{12} n^{-2}\phidot^2+\frac{1}{12} b^{-2}\phi'{}^2-\frac{1}{4}\alpha^2 e^{-2\phi} \el
  \frac{n'}{n}\frac{\adot}{a}+\frac{a'}{a}\frac{\bdot}{b} -\frac{\adot'}{a} &=& \frac{1}{6}\phidot\, \phi' \ .
\eea

Taking $y=y_1=0$, using \re{branematter} and taking into account $n(t,0)=1$,
we have
\bea
  -\frac{1}{36 M_5^6}\rho_{b(i)}(\rho_{b(1)} + 3 p_{b(1)})-\left(\frac{\addot_0}{a_0} + \frac{\adot_0^2}{a_0^2}\right)
&=& \frac{1}{12} \phidot_0^2 + \frac{1}{12} b_0^{-2}\phi_0{}'{}^2 - \frac{1}{4}\alpha^2 e^{-2\phi_0} \el
  \rhodot_{b(1)} + 3 \frac{\adot_0}{a_0} (\rho_{b(1)} + p_{b(1)})
&=& M_5^3 b_0^{-1} \phidot_0\, \phi'{}_{\!\! 0} \ ,
\eea

\noindent where $a_0(t)\equiv a(t,0)$ and so on. Expressing $\phi'{}_{\!\! 0}$
in terms of $\phi_0$ by using the equation of motion of $\phi$ on the branes, 
\re{eomphib}, we have
\bea
  \label{branefr}  \left(\frac{\addot_0}{a_0} + \frac{\adot_0^2}{a_0^2}\right) + \frac{1}{36 M_5^6}\rho_{b(1)}(\rho_{b(1)} + 3 p_{b(1)}) + \frac{1}{12} \phidot_0^2 + \frac{1}{2}\alpha^2 e^{-2\phi_0} &=& 0 \\
  \label{branecons1} \rhodot_{b(1)} + 3 \frac{\adot_0}{a_0} (\rho_{b(1)} + p_{b(1)}) + 3\,M_5^3\alpha_1\, e^{-\phi_0} \phidot_0 &=& 0 \ .
\eea

Multiplying \re{branefr} by $2\adot_0/a_0$ and \re{branecons1} by
$2\rho_{b(1)}/36$ and subtracting, and expressing $\rho_{b(1)}$ and $p_{b(1)}$
in terms of $\rho_m$ and $p_m$ as given by \re{rhoandp}, we get
\bea
  \label{prehubble} \pat_t\left( \frac{\adot_0^2}{a_0^2} - \frac{1}{36 M_5^6}\rho_{b(1)}^2 + \frac{1}{4}\alpha^2 e^{-2\phi_0} \right) + 4 \frac{\adot_0}{a_0}\left(\frac{\adot_0^2}{a_0^2}-\frac{1}{36 M_5^6}\rho_{b(1)}^2 + \frac{1}{4}\alpha^2 e^{-2\phi_0} \right) \el
  + \frac{1}{6}\frac{\adot_0}{a_0}\phidot_0^2 - \frac{1}{6 M_5^3}\alpha_1\, e^{-\phi_0} \phidot_0 \rho_m &=& 0 \\
  \label{branecons2} \rhodot_m + 3 \frac{\adot_0}{a_0} (\rho_m + p_m) &=& 0 \ .
\eea

Integrating \re{prehubble}, we obtain the Hubble law on the visible brane:
\bea
  \label{hubble1} \frac{\adot_0^2}{a_0^2} &=& \frac{1}{36 M_5^6}\rho_{b(1)}^2 - \frac{1}{4}\alpha^2 e^{-2\phi_0} + \frac{{\cal C}}{a_0^4} + f(t) \ ,
\eea

\noindent where $\cal C$ is a constant\footnote{For the bulk origin and
interpretation of this term see \cite{Mohapatra:2000, Ida:1999,
Mukohyama:2000}.} and $f$ is a solution of the following equation:
\bea \label{f1}
  \fdot + 4 \frac{\adot_0}{a_0} f &=& \frac{1}{6 M_5^3}\alpha_1\, e^{-\phi_0} \phidot_0 \rho_m - \frac{1}{6}\frac{\adot_0}{a_0}\phidot_0^2 \ .
\eea

\noindent Note that when $\phidot_0=0$, $f$ reduces to the $\cal C$-term.
The Hubble law \re{hubble1} is a special case of the Hubble law with a
bulk scalar field with a general potential considered in \cite{Brax:2001}.

The Hubble law \re{hubble1} has a quadratic dependence on the total
energy density of the visible brane. So the issue is whether we can
recover the standard linear dependence on the energy density of brane matter.

Let us first integrate the equation \re{f1}. We write $\rho_m=\rho_r+\rho_d$,
to separate the contributions of radiation energy density $\rho_r$
$(\propto a_0^{-4})$ and energy density of other brane matter $\rho_d$
({\em e.g.} dust). The result for $f$ is
\bea \label{f2}
  f &=& \frac{1}{6 M_5^3} a_0^{-4} \int\!\! dt\, a_0^4\,\alpha_1\, e^{-\phi_0}\phidot_0\rho_m - \frac{1}{6} a_0^{-4} \int\!\! dt\, a_0^3\adot_0\phidot_0^2 \el
  &=& -\frac{1}{6 M_5^3}\alpha_1 (e^{-\phi_0}-e^{-\phi_0(t_0)})\rho_r
+ \frac{1}{6 M_5^3} a_0^{-4} \int\!\! dt\, a_0^4\,\alpha_1\, e^{-\phi_0}\phidot_0\rho_d \el
  & & - \frac{1}{6} a_0^{-4} \int\!\! dt\, a_0^3\adot_0\phidot_0^2 \ .
\eea

Then, we substitute in \re{hubble1} the brane energy density $\rho_{b(1)}$
in terms of $\rho_m$ as given by \re{rhoandp}. Combining \re{branecons2},
\re{hubble1} and \re{f2}, we obtain the ekpyrotic generalization of the
Friedmann-Robertson-Walker equations on the visible brane:
\bea
  \label{ekpyroFRW1} & & \frac{\adot_0^2}{a_0^2} \,=\, \frac{1}{6 M_5^3}\alpha_1 e^{-\phi_0(t_0)}\rho_r+\frac{1}{6 M_5^3}\alpha_1 e^{-\phi_0}\rho_d + \frac{1}{36 M_5^6}\rho_m^2 + \frac{{\cal C}}{a_0^4} \el
  & &\qquad\ \, + \frac{1}{6 M_5^3} a_0^{-4} \int\!\! dt\, a_0^4\,\alpha_1\, e^{-\phi_0}\phidot_0\rho_d - \frac{1}{6} a_0^{-4} \int\!\! dt\, a_0^3\adot_0\phidot_0^2 \\
  \label{ekpyroFRW2} & & \rhodot_m + 3 \frac{\adot_0}{a_0} (\rho_m + p_m) \,=\, 0 \ .
\eea

Note that the bulk and brane terms proportional to $\alpha^2 e^{-2\phi_0}$
have cancelled. So the effective cosmological constant equals zero. We will
comment on this after considering how to obtain the rest of standard cosmology.

\paragraph{Towards standard cosmology.}

The brane equations \re{ekpyroFRW1} and \re{ekpyroFRW2} are to be
compared with the standard four-dimensional FRW equations:
\bea
  \label{FRW1} & & \frac{\adot^2}{a^2} \,=\, \frac{8\pi}{3}G_N\rho_m \\
  \label{FRW2} & & \rhodot_m + 3 \frac{\adot}{a} (\rho_m + p_m) \,=\, 0 \ ,
\eea

\noindent where $G_N=1/(8\pi M_4^2)$ is the four-dimensional Newton's constant.

A comparison of \re{ekpyroFRW2} and \re{FRW2} shows that in the ekpyrotic
scenario the brane energy-momentum tensor satisfies the standard
conservation law. This is because the energy lost (or gained) via
time-dependence of the term $e^{-\phi_0}$ on the brane is exactly compensated
by the (continuous part of the) energy flow $T_{ty}$.

Since space terminates at the branes and
$T_{ty}=\frac{1}{2}\phidot\,\phi'$, $\phidot_0=0$ seems like a natural
boundary condition. However, it is not clear to us whether this boundary
condition is absolutely necessary. In what follows, we will keep
$\phidot_0\neq0$ (and see that things simplify considerably if $\phidot_0=0$).

A comparison of \re{ekpyroFRW1} and \re{FRW1} shows that
on the visible brane, the standard Hubble law and thus standard
cosmology are recovered to a good approximation, as long as the conditions
\bea
  \label{planck} G_N &=& \frac{\alpha_1}{16\pi M_5^3} e^{-\phi_0} \\
  \rho_m\frac{M_4^2}{96\pi M_5^6} &\ll& 1 \\
  \phidot_0 &\approx& 0, \quad  e^{\phi_0(t)} \approx e^{\phi_0(t_0)} \
\eea

\noindent are satisfied.

In addition, we should also have ${\cal C}/a^4\ll \rho_m/M_4^2$.
Let us interpret these conditions. First, in order to obtain standard
gravity with a positive Newton's constant, the visible brane
must have positive tension, $\alpha_1=-\alpha>0$, in contrast to the negative
tension used in \cite{Khoury:2001a}. This is our second result. A similar
result is well known in the context of the Randall-Sundrum model
\cite{Csaki:1999a, Cline:1999, Binetruy:1999b} (for a systematic account see
\cite{Shiromizu:1999}); in the context of the Ho\v{r}ava-Witten model this
was discussed in \cite{Mennim:2000}. Here, our goals were to i) provide an
explicit derivation of the Hubble law in the context of the
ekpyrotic scenario, and ii) demonstrate how the issue of obtaining
standard cosmology is related to the brane tension assignment.

The second condition is required to ensure that the term quadratic in $\rho_m$
is subleading with respect to the term linear in $\rho_m$. If $M_4\sim M_5$,
as is natural, this condition is easily satisfied: the quadratic term is
vanishingly small and has no observable consequences.

The third condition is required for two reasons: first, to have Newton's
constant to approximately constant on the brane (or indeed, to be able to
define Newton's constant at all), and second, to have the term involving
$\phidot_0^2$ be small.

Note the novel feature of \re{ekpyroFRW1} that the gravitational coupling of
radiation is different from the coupling of other types of matter. The
time-dependence of $\phi_0$ which changes Newton's constant has no effect on
the coupling of radiation. This point is also illustrated by the
$yy$--component of the Einstein equation, \re{branefr}. Expressing
$\rho_{b(1)}$ and $p_{b(1)}$ in terms of $\rho_m$ and $p_m$ as given by
\re{rhoandp}, \re{branefr} reads
\bea
  \frac{\addot_0}{a_0} + \frac{\adot_0^2}{a_0^2}
= \frac{1}{12 M_5^3}\alpha_1 e^{-\phi_0}(\rho_m-3 p_m) - \frac{1}{36 M_5^6}\rho_m(\rho_m + 3 p_m) - \frac{1}{12} \phidot_0^2 \ .
\eea

We find again that radiation does not feel the time-dependence of $\phi_0$.
During the radiation dominated era of cosmology, a time-varying
$\phi_0$ will only produce deviations from standard cosmology via the
additional source proportional to $\phidot_0^2$. So, we might safely allow a
time-dependent $\phi_0$ without significantly affecting for example
nucleosynthesis. All that is required is $\phidot_0^2<\adot_0^2/a_0^2$ by a
couple of orders of magnitude. Of course, when the equation of state of
brane matter begins to significantly deviate from that of radiation,
$\phidot_0$ should be rather small to allow standard cosmology.

A comparison of big bang nucleosynthesis and the Newton's law observed
today shows that the gravitational coupling of radiation at the time of
nucleosynthesis is close to the gravitational coupling of 
non-relativistic matter observed today \cite{Accetta:1990}.
This means that either the coupling of radiation and the coupling of
other matter have been almost the same since nucleosynthesis (requiring
$\phidot_0=0$ to a good accuracy; see \cite{Guenther:1998} for
experimental limits on $\phidot_0$) or the gravitional coupling of
radiation is an attractor for the general coupling. The first
possibility seems more likely.

By performing a partial integration and using \re{ekpyroFRW2}, we can write
\re{ekpyroFRW1} as
\bea \label{ekpyroFRW1b}
  \frac{\adot_0^2}{a_0^2} &=& \frac{1}{6 M_5^3}\alpha_1 e^{-\phi_0(t_0)}\rho_r + \frac{1}{36 M_5^6}\rho_m^2 + \frac{{\cal C}}{a_0^4} \el
  & & +\, a_0^{-4} \int\!\! dt\, a_0^3\adot_0\left(\frac{1}{6 M_5^3}\alpha_1\,e^{-\phi_0} (\rho_d - 3 p_d)-\frac{1}{6}\phidot_0^2\right) \ .
\eea

As the universe expands, the contributions from the brane energy density and
pressure (as well as the $\cal C$-term) approach zero, assuming that
$\phi_0$ does not decrease so rapidly that $e^{-\phi_0} \rho_m$ grows.
Since the \emph{l.h.s.} of \re{ekpyroFRW1b} is positive
definite and the term involving $\phidot_0^2$ is negative definite, we see
that $\phidot_0$ approaches zero as the universe becomes asymptotically
empty. This is in agreement with our observation in section 2.2 that the
magnitude of $\phidot$ (if different from zero) is related to the
magnitude of $\rho_m$.

\paragraph{Comments.}

As in the well known case of the Randall-Sundrum model,
the cosmology on the branes of the ekpyrotic model differs from FRW cosmology
in two ways: the square of the Hubble parameter is proportional to the square
of the brane energy density and there is a contribution due to bulk curvature.
The usual linear dependence is obtained by having a positive cosmological
constant on the brane, at the price of having a term involving
the cosmological constant squared.\footnote{For an alternative mechanism, see
\cite{Kanti}.} It is notable that in the ekpyrotic model, this term is exactly
cancelled by a contribution from the bulk curvature, leaving a vanishing
effective cosmological constant on the brane. See also \cite{Mennim:2000}.

This story is familiar from the Randall-Sundrum model. For cosmological
solutions with a cosmological constant as the only bulk source,
only one particular relation between the bulk and brane cosmological constants
gives a vanishing effective cosmological constant on the brane.
This apparent fine-tuning has its origin in the BPS conditions \cite{Cvetic}.
This is also the case in the ekpyrotic model: the relation between the brane
tension and charge that characterises BPS solutions also cancels the effective
cosmological constant on the brane for cosmological solutions. A different
prefactor of ${\cal A}_{ABCD}$ in \re{action1} would yield a non-zero effective
cosmological constant on the brane.

The behaviour of terms involving bulk curvature that do not cancel against the
brane tension, namely the $\cal C$-term and $\phidot_0$-terms, as well as
the value of the gravitational coupling of radiation, can only be provided by
the  bulk equations. If one imposes the seemingly natural
boundary condition $\phidot_0=0$, then the only ambiguity is related to the
$\cal C$-term and (equivalently) to the gravitational coupling of radiation.
Then the ekpyrotic brane cosmology looks, in the homogeneous and isotropic
approximation, the same as brane cosmology in the Randall-Sundrum model
with a cosmological constant as the only bulk source. However, it is
not clear whether the condition $\phidot_0=0$ can be imposed: we have observed
that it is possible to obtain standard cosmology if $\phidot_0$ is small,
but not shown that it actually is small.
In section 2.1 we saw that in order to have time-dependent brane matter
we must have either $\bdot\neq0$ or $\phidot\neq0$, or both. 
Note that, as in the Randall-Sundrum model, the
time-dependence of $b$ does not directly enter into the brane expansion
equations, so that from the point of view of brane cosmology there is no
reason to exclude even a radical time-dependence in $b$ (as opposed to $\phi$).
For these reasons it would be interesting to find an explicit bulk
solution for the post-ekpyrosis universe, possibly as a perturbation
around the BPS solution.

\section{Discussion}

We have two main results. We have observed that the visible brane must have
positive tension in order to support approximately standard gravity and shown
that time-dependent brane matter will cause the size of the hidden dimensions
to change in time.

In the original example of a realization of the ekpyrotic scenario
\cite{Khoury:2001a} (version 1) it was implied that the visible brane
should, in that setting, have negative tension. In \cite{Kallosh:2001} it was
argued that the visible brane cannot have negative tension and a
variant with the signs of the brane tensions flipped was investigated. The
authors of \cite{Khoury:2001a} responded that the visible brane could in
principle have either positive or negative tension \cite{Khoury:2001b}, and
that in particular the original setting allows for any sign for the tension of
the visible brane \cite{Khoury:2001a} (version 2). Explicit constructions with
negative tension visible branes were also presented \cite{Donagi:2001}. The
debate continued in \cite{KKLT}, where it was argued that obtaining standard
Ho\v{r}ava-Witten phenomenology requires the visible brane to have positive
tension.

Now, it is again argued that the visible brane should have positive tension.
However, we emphasize that our argument follows a different path than the
previous investigations. We do not study which tension assignments
are possible in M-theory or what is the resulting heterotic phenomenology.
Rather, we consider the viewpoint of observers on the visible brane and
study the gravity seen by them.

It is worth mentioning that the four-dimensional Newton's
constant given by the induced Einstein equation on the brane, \re{planck},
differs from the one obtained in the usual dimensional reduction
\cite{Witten:1996}: the inverse of the size of the fifth dimension
$\pi R$ is replaced by the brane tension $\alpha_1$.

It is perhaps fortunate that the size of the fifth dimension does not appear
in the Newton's constant measured on the brane. We have seen that
time-dependent brane matter will induce time-dependence in the
size of the fifth dimension or in the size of the Calabi-Yau threefold.
Of course, we have studied only the simplified action used in
\cite{Khoury:2001a}. The five-dimensional action of heterotic M-theory is
much more complicated and can include a plethora of other moduli
\cite{Lukas:1998a, Lukas:1998b}. It remains to be seen whether this effect
is a general feature of the ekpyrotic scenario, present in more complicated
settings as well.

By deriving the Hubble law on the visible brane, we have shown
that it is possible to obtain approximately standard cosmology
(provided that the brane has positive tension). In contrast to the size of
the fifth dimension, the size of the Calabi-Yau threefold appears directly
in the Hubble law. Its time-dependence may have observable
cosmological effects. In particular, Newton's constant may be
time-dependent, though curiously the gravitational coupling of
radiation is always time-independent. In order to estimate the magnitude of
the effects arising from the time-dependence of the Calabi-Yau threefold, it
would be interesting to find an explicit solution where one could see the
relation between brane matter and the size of the hidden dimensions.

\section*{Acknowledgements}

K.E. has been supported in part by the Academy of Finland under the contract
35224. We thank Vijay Balasubramanian for discussions. E.K-V. would like to
thank the University of Pennsylvania for hospitality at the inception of this
work. We also thank Anne Davis and Andrei Linde for comments.

\end{document}